\documentclass[conference, onecolumn]{IEEEtran}
\IEEEoverridecommandlockouts

\usepackage{cite}

\usepackage[utf8]{inputenc} 
\usepackage{url}

\usepackage{graphicx}
\usepackage{textcomp}
\usepackage{algorithm}
\usepackage{algpseudocode}
\usepackage[cmex10]{amsmath}
\usepackage[T1]{fontenc}
\usepackage{amssymb}
\usepackage{caption}
\usepackage{subcaption}
\usepackage{amsthm}
\usepackage{enumitem}
\usepackage{algorithmicx}
\usepackage{algpseudocode}
\newtheorem{conjecture}{Conjecture}

\theoremstyle{definition}

\usepackage{listings}
\usepackage{mathtools}
\usepackage[dvipsnames]{xcolor}
\usepackage{tabularx}
\newcolumntype{L}{>{\centering \arraybackslash}X}

\definecolor{keywordcolor}{rgb}{0.0,0.0,0.6}
\definecolor{commentcolor}{rgb}{0.0,0.5,0.0}
\definecolor{stringcolor}{rgb}{0.58,0,0.82}

\def\BibTeX{{\rm B\kern-.05em{\sc i\kern-.025em b}\kern-.08em
    T\kern-.1667em\lower.7ex\hbox{E}\kern-.125emX}}

\newcommand{\un}[1]{\underline{#1}}

\newtheorem{defn}{Definition}
\newtheorem{thm}{{\cal T}heorem}
\newtheorem{cor}{Corollary}
\newtheorem{prop}{Proposition}
\newtheorem{lem}{Lemma}
\newtheorem{conj}{Conjecture}
\newtheorem{constr}{Construction}
\newtheorem{note}{Remark}

\newtheorem{example}{Example}
\newcommand{\bit}{\begin{itemize}}
	\newcommand{\eit}{\end{itemize}}
\newcommand{\bcor}{\begin{cor}}
	\newcommand{\ecor}{\end{cor}}
\newcommand{\beq}{\begin{equation}}
	\newcommand{\eeq}{\end{equation}}
\newcommand{\beqn}{\begin{equation}}
	\newcommand{\eeqn}{\end{equation}}
\newcommand{\bea}{\begin{eqnarray}}
	\newcommand{\eea}{\end{eqnarray}}
\newcommand{\bean}{\begin{eqnarray*}}
	\newcommand{\eean}{\end{eqnarray*}}
\newcommand{\ben}{\begin{enumerate}}
	\newcommand{\een}{\end{enumerate}}
\newcommand{\bdefn}{\begin{defn}}
	\newcommand{\edefn}{\end{defn}}
\newcommand{\bnote}{\begin{note}}
	\newcommand{\enote}{\end{note}}
\newcommand{\bprop}{\begin{prop}}
	\newcommand{\eprop}{\end{prop}}
\newcommand{\blem}{\begin{lem}}
	\newcommand{\elem}{\end{lem}}
\newcommand{\bthm}{\begin{thm}}
	\newcommand{\ethm}{\end{thm}}
\newcommand{\bconj}{\begin{conj}}
	\newcommand{\econj}{\end{conj}}
\newcommand{\bconstr}{\begin{constr}}
	\newcommand{\econstr}{\end{constr}}
\newcommand{\bpf}{\begin{proof}}
	\newcommand{\epf}{\end{proof}}
\newcommand{\bprf}{{\em Proof: }}
\newcommand{\eprf}{\hfill $\Box$}

\newcommand{\cX}{\mathcal{X}}
\newcommand{\cY}{\mathcal{Y}}
\newcommand{\cS}{\mathcal{S}}
\newcommand{\Wd}{\mathcal{W}_{\delta}}
\newcommand{\Pd}{\mathcal{P}_{\delta}}

\begin{document}

\title{Extreme Points of the $(0,\delta)$-LDP Polytope with Small Input Size and Arbitrary Output Sizes}

\author{\IEEEauthorblockN{ Supriya Rawat}
\IEEEauthorblockA{\textit{CSE, IIT Hyderabad} \\ supriya.rawat.1201@gmail.com} 
\and
\IEEEauthorblockN{Myna Vajha}
\IEEEauthorblockA{\textit{EE, IIT Hyderabad} \\
mynav@ee.iith.ac.in}
\and
\IEEEauthorblockN{ Gowtham R. Kurri}
\IEEEauthorblockA{\textit{IIIT Hyderabad} \\ gowtham.kurri@iiit.ac.in}
\and
\IEEEauthorblockN{ Anand Sarwate}
\IEEEauthorblockA{\textit{ECE, Rutgers University} \\
anand.sarwate@rutgers.edu}
}

\maketitle
 \begin{abstract}
The structure of locally differentially private (LDP) mechanisms can be understood through the geometry of the corresponding privacy polytope. While the extreme points of the \( (\epsilon,0)\)-LDP polytope are well characterized (Kairouz {et al.}, 2014; Holohan {et al.}, 2017), comparatively little is known for the \((\epsilon,\delta)\)-LDP polytope with \(\delta>0\). Recent work (Elangovan and Jog, 2024) has shown that even in the special case \(\epsilon=0\), the \( (0,\delta) \)-LDP privacy polytope exhibits fundamentally different behaviour. In this work, we provide complete characterizations of the extreme points for the low-input-alphabet regime \(k=2\) and \(k=3\) and with arbitrary output alphabet size \(m \). We also identify new extreme mechanisms for larger input alphabet sizes \(k\), of the star configuration type, as introduced by Elangovan and Jog (2024).    
\end{abstract}



\section{Introduction}
Local differential privacy (LDP)~\cite{kasiviswanathan2011can,duchi2013local} has emerged as a central privacy notion in distributed statistical inference~\cite{duchi2018minimax}, learning~\cite{DuchiJW14}, and hypothesis testing~\cite{AcharyaCT20}. LDP requires each individual entry in a database to be randomized before any aggregation, making it particularly relevant in large-scale and decentralized systems. This is in contrast to centralized differential privacy (DP)~\cite{dwork2006differential}, where database entries are collected by a trusted curator, and privacy is enforced only at the level of the released aggregate. From an information-theoretic perspective, LDP constraints significantly limit what can be learned from the data, creating a clear tradeoff between privacy and utility.

In hypothesis testing and estimation problems under LDP~\cite{PensiaAJL25}, optimal mechanisms are characterized as those that maximize appropriate divergence measures between the induced output distributions, which are directly related to sample complexities and error exponents. Owing to the convexity of these divergence measures, such as the Hellinger distance, in their arguments, it suffices to restrict attention to the extreme points of the convex polytope of LDP mechanisms~\cite{PensiaAJL25}. Consequently, the study of these extreme points is key to solving such optimization problems.

Kairouz \emph{et al.}~\cite{kairouz2014extremal}  introduced the \emph{extremal}
(staircase) mechanisms, a finite subset of the $(\epsilon,0)$-LDP polytope and showed they contain an optimizer of any sub-linear utility, attained with $m\le k$. Holohan \emph{et al.}~\cite{holohan2017extreme} characterized all extreme points with $1$, $2$, or $k$ non-zero columns.

In contrast to the $(\epsilon, 0)$-LDP case, the structure of the privacy polytope under $(\epsilon,\delta)$-LDP with $\delta>0$ is far less understood. Recent work~\cite{extremeLDP} shows that even in the simplest case $\epsilon=0$, the geometry of the $(0,\delta)$-privacy polytope exhibits a fundamentally different behaviour. In particular, while all extreme points can be fully characterized for output alphabet sizes $m\leq 3$, explicit examples for $m\geq 4$ reveal some extreme points with output alphabet size exceeding the input alphabet size, i.e., $m>k$.  
\subsection{Our Contributions}
We study the extreme points of the $(0,\delta)$-LDP polytope, with our main
results concerning the low-input-alphabet regime $k\in\{2,3\}$ and arbitrary output alphabet size $m$. (a) For arbitrary $k,m\in\mathbb{N}$, we completely characterize the extreme configurations containing a \emph{singleton row}, that is, a row with a single $1$ and all remaining entries $0$ (Lemma~\ref{lem:singletonextreme}). (b) We prove the tight-point conjecture
of~\cite{extremeLDP} for $k=3$ (Lemma~\ref{lem:tightpoint}). (c) Building on (a) and (b), we obtain complete characterizations of the extreme points for $k=2$ and $k=3$ (Theorems~\ref{thm:k=2case} and~\ref{thm:k3}). (d) We also
identify new extreme mechanisms for larger input alphabet sizes, extending the class of star configurations recently introduced by Elangovan and
Jog~\cite{extremeLDP} (Lemmas~\ref{lem:starconfig0}, \ref{lem:starconfig1}
and~\ref{lem:starconfig2}).

\subsection{Notation} Finite alphabets will be denoted by calligraphic type, e.g. $\cX, \cY$. For a positive integers $a, b$ with $a < b$,  $[a] = \{1,2,\ldots,a\}$ and $[a:b] = \{a, a+1, \ldots, b\}$. 

\section{Background}
Let $\cX = [k]$ and $\cY = [m]$ denote the input and output alphabets. A privacy mechanism is a randomized map/channel $M : \cX \to \cY$. We can represent a mechanism by a $k\times m$ row-stochastic matrix $T \in [0,1]^{k\times m}$ where the $x$-th row, denoted by $T_x$, is the conditional distribution on $\cY$ given an input $x \in \cX$, so $T_{x,y} = \Pr[M(x) = y]$.

\bdefn[$(\epsilon,\delta)$-Local Differential Privacy]
A privacy mechanism satisfies $(\epsilon,\delta)$-Local Differential Privacy (LDP) if for all pairs of inputs 
$x, x' \in \mathcal{X}$ and for all measurable subsets 
$\cS \subseteq \mathcal{Y}$,
\bean
\Pr[M(x) \in \cS] \;\leq\; e^{\epsilon} \, \Pr[M(x') \in \cS] + \delta.
\eean
\edefn
Let $\Pd$ denote the collection of stochastic matrices that satisfy the $(0,\delta)$-LDP constraint given by:
\begin{align}
\Pd =\Bigl\{
 &T \in \mathbb{R}^{k \times m} \,\Big|\,
T_{x,y} \ge 0,\;
\sum\nolimits_{y=1}^{m} T_{x,y} = 1 \ \forall x \in [k],\nonumber\\
&d_{\mathrm{TV}}(T_x, T_{x'}) \le \delta,
\ \forall x,x' \in [k]
\Bigr\},\label{eq:ldp_polytope}
\end{align}\label{eq:ldp_polytope}
where $d_{TV}(\cdot,\cdot)$ denotes the total variation (TV) distance. The set $\Pd$ is a convex polytope, as the condition on total variation distance between any two rows can be expressed as a set of linear inequalities.

\bdefn[Extreme matrices of $\Pd$]
An \emph{extreme point} of $\Pd$ is a matrix $T \in \Pd$ that cannot be written as a non-trivial convex combination of two other distinct matrices in $\Pd$. In other words,
whenever
\[
T = \lambda T^{(1)} + (1-\lambda) T^{(2)}, \quad 0 < \lambda < 1, \quad T^{(1)}, T^{(2)} \in \Pd,
\]
we must have $T^{(1)} = T^{(2)} = T$.
\edefn

Because the ordering of the input and output alphabets is arbitrary, if $T$ is an extreme matrix, permuting the rows and/or columns of $T$ will also yield an extreme matrix.

\bdefn[Configuration and extremal configuration]
For a row-stochastic matrix $T \in \Pd$, the associated \emph{configuration} $\mathcal{C}(T)$ is defined as the set of distinct row
vectors of $T$, i.e., repeated rows are removed from $T$ to get $C(T)$.

The elements of a configuration are called \emph{points} and are viewed as points in the $(m-1)$-dimensional probability simplex in $\mathbb{R}^m$. A configuration is called \emph{extremal} if it arises as the configuration associated with some extreme matrix $T \in \Pd$.
\edefn

In this section we characterize the extreme matrices of $\Pd$ for $k=2$ and
$k=3$ with arbitrary output alphabet size $m$. Our main tool is
\cite[Theorem~4]{extremeLDP}, restated below, which gives the conditions a
perturbation matrix $E$ must satisfy in order that $T \pm E \in \Pd$ for a
given $T \in \Pd$. This results in a direct test for extremality: $T$ is an
extreme point of $\Pd$ if and only if the all-zero matrix is the only such
$E$. We write $\epsilon_{i,j}$ for the $(i,j)$ entry of $E$; throughout the
paper, whenever a perturbation matrix is specified, any entry not given
explicitly is assumed to be zero.

\bthm[Perturbation Equations~\cite{extremeLDP}] \label{thm:perturb}
Let $T \in \Pd$,
and let $E$ be a matrix of the same dimensions as $T$ with 
$\|E\|_{\infty}$ small. Then $T \pm E $ are both 
elements of $\Pd$ if and only if 
$E$ satisfies the following constraints:
\begin{enumerate}
    \item $\displaystyle \sum_{j=1}^m \epsilon_{ij} = 0, \quad \forall i \in [k].$
    \item $\epsilon_{ij} = 0$ whenever $T_{ij} = 0$.
    \item $\epsilon_{ij} = \epsilon_{kj}$ whenever 
          $d_{\mathrm{TV}}(T_i, T_k) = \delta$ and $T_{ij} = T_{kj}$.
    \item $\displaystyle \sum_{j=1}^m \operatorname{sgn}(T_{ij} - T_{kj})(\epsilon_{ij} - \epsilon_{kj}) = 0$
          whenever $d_{\mathrm{TV}}(T_i, T_k) = \delta$.
\end{enumerate}
\ethm

\blem[Localisation of Extreme Points~\cite{extremeLDP}]
\label{lem:localisation}
Let $T_1,\ldots,T_k$ form an extreme configuration. Then there exists a unique coordinate $j_0 \in [m]$ such that $T_{ij} \le \delta$ for all $i\in[k]$ and all $j \in [m]\setminus\{j_0\}$.
The configuration is said to be \emph{localised} at the coordinate $j_0$.
\elem

\section{Extreme Matrices for $k=2, 3$}
\bdefn[Tight Point of Configuration]
A \emph{point of a configuration} is \emph{tight} if it is the maximal distance $\delta$ away from every other point.
\edefn

\begin{conjecture}[\!\!{\cite{extremeLDP}}]
There exists a tight point for all extreme configurations.
\end{conjecture}  
We prove that this conjecture is true for the $k=3$ case and use this property to characterise the extreme matrices for the $k=3$ case in Theorem~\ref{thm:k3}. We start by showing that extreme configuration that contains the singleton row (a row with one 1 and all other entries 0) as a point is such that every point is tight w.r.t every other point.

\blem\label{lem:singletonextreme}
For any $k,m\in\mathbb{N}$, the only possible extreme configuration with a singleton-row as an element is as shown below up to row/column permutations.
\bea
\label{eq:extremesingletonrow}
T=\left[\begin{array}{ccccccc}
     1 & 0 & 0 & \cdots & 0 & \cdots & 0 \\
     1-\delta & \delta & 0 & \cdots & 0 & \cdots & 0 \\
     1-\delta & 0 & \delta & \cdots & 0 & \cdots & 0 \\
     \vdots & \vdots & \vdots & \ddots & \vdots & \ddots & \vdots \\
     1-\delta & 0 & 0 & \cdots & \delta & \cdots & 0
\end{array}\right].
\eea
\elem
We  present a decomposition Lemma that will be used to prove Lemma~\ref{lem:singletonextreme}. 

\bdefn\label{defn:Wd}
We define,
\[
  \Wd=\bigl\{\, T \in \mathcal{S} \mid T_{i,1} \ge 1-\delta \text{ for all } i \in [k] \,\bigr\}.
\]
where $\mathcal{S}$ is the collection of $k \times m$ stochastic matrices.
\edefn

If $T \in W_{\delta}$ the for all $i \in [k]$ it follows that the off-mass defined by $s_i \triangleq \sum_{j=2}^m T_{i,j}$ is at-most $\delta$.

\blem $W_{\delta} \subseteq \Pd$
\elem
\bprf
Let $T\in \Wd$. It can be seen that:
\bean
d_{TV}(T_i, T_{\hat{i}}) &=& (T_{i,1}-T_{\hat{i},1})_{+} + \sum_{j=2}^{m} (T_{i,j}-T_{\hat{i},j})_{+}\\
&\le& (s_{\hat{i}}-s_i)_{+} + s_i \le \delta.
\eean
Here $s_i = \sum_{j=2}^m T_{i,j}$ and $T_{i,1}-T_{\hat{i},1}=(1-s_i)-(1-s_{\hat{i}})=s_{\hat{i}}-s_i$. The last inequality follows as $s_i, s_{\hat{i}} \le \delta$. Therefore, $T \in \Pd$.
\eprf


\bean
\text{Let }R_0 &=& [1~0~\cdots~0]\\
R_u &=& [1-\delta, 0,~\dots, \delta \text{ (at col } u), \dots, 0].
\eean
for any $u \in [2:m]$. Using these rows, we define the matrices:
\bean
E(u_1, u_2, \cdots, u_k)=\begin{bmatrix}
    R_{u_1}^\top & R_{u_2}^\top & \cdots & R_{u_k}^\top
\end{bmatrix}^{\top}.
\eean

\blem [Decomposition Lemma]\label{lem:decom}
Let $T \in \Wd$, then $T$ is a convex combination of the matrices in the set:
\bean
\{E(u_1, \cdots, u_k) \mid u_i \in \{0\} \cup [2:m],   ~ \forall  i \in [k]\}.
\eean
Consequently, if $T \in \Wd$ is an extreme point of $\Pd$ then $T = E(u_1, \cdots, u_k)$ for some $u_1, \cdots, u_k \in \{0\} \cup [2:m]$.
\elem
\begin{proof}
Let $T \in \Wd$ for all $i \in [k]$, $T_{i,1} \ge (1-\delta)$ and therefore it can be expressed as:
\bean
T_i = p_0^{(i)} R_0 + \sum_{u=2}^m p_u^{(i)} R_u
\eean
where  $p_u^{(i)}=\frac{T_{i,u}}{\delta}$ and $p_0^{(i)} =  1-\sum_{u=2}^m p_u^{(i)}$.  We now show that the matrix $T$ can be represented as convex combination of matrices in $\{E(u_1, \cdots, u_k) \mid u_i \in \{0\} \cup [2:m]\}$ as shown below:
\bean
  T = \sum_{u_1, \cdots, u_k } \Bigl(  \prod_{i=1}^k p^{(i)}_{u_i}    \Bigr) E(u_1, \cdots, u_k).
\eean
To see this the first row of the RHS above is of the form:
\bean
\sum_{u_1, \cdots, u_k } \Bigl(  \prod_{i=1}^k p^{(i)}_{u_i}    \Bigr) R(u_1) = \Bigl( \prod_{i=2}^k \Bigl(\sum_{u_i}  p^{(i)}_{u_i}\Bigr) \Bigr) \Bigl(\sum_{u_1} p^{(1)}_{u_1} R_u\Bigr).
\eean
It is clear to that this is same as $T_1$. Similarly it can be verified that this is true for every row.
\end{proof}

\textit{Proof of Lemma~\ref{lem:singletonextreme}}:
Let $T$ be an extreme configuration, $T_i$ denote the $i$-th row of $T$, and assume the first row $T_1=[1~0~0~\cdots~0]$ is a singleton row. Since $T \in \Pd$,  it follows that the total variation distance between $T_1$ and any other row $T_i$ must satisfy $d_{TV}(T_1, T_i) \le \delta$. 
For any row $i \ge 2$:
\bean
d_{TV}(T_1, T_i) &=& \frac{1}{2} \left( (1 - T_{i,1}) + \sum_{j>1} T_{i,j} \right) \\
&=& 1 - T_{i,1} \le \delta \implies T_{i,1} \ge 1-\delta
\eean
This implies that $T \in \Wd$ and therefore by the decomposition lemma~\ref{lem:decom} it follows that $T$ has to be of the form shown in equation~\eqref{eq:extremesingletonrow} upto row/column permutations as $T$ has no repeating rows.
\eprf

\blem \label{lem:deltarow}
Let $T \in {\cal P}_{\delta}$ be an extreme configuration. For every row $T_i$ of $T$ 
there exists a row $j_i \in [k] \setminus \{i\}$ such that $d_{TV}(T_i, T_{j_i}) = \delta$.
\elem
\bprf
For the case when $T_i$ is a singleton row, from Lemma~\ref{lem:singletonextreme} it follows that every other row is tight w.r.t $i$-th row and the statement follows. Suppose $T_i$ is not a singleton row and suppose there does not exist a row that is tight w.r.t it, then $d_i = \max_{j \in [k] \setminus \{i\} }d_{TV}(T_i, T_j) < \delta$. Let $S = \{j \mid T_{i,j} > 0\}$. Since $T_i$ is not a singleton row, we have $|S|\ge 2$ and let $j_1, j_2 \in S$. Let  $\epsilon^* = \min (T_{i, j_1}, T_{i, j_2}, \delta-d_i) $ and let the perturbation matrix $E$ be defined such that $\epsilon_{i,j_1} = \epsilon^* = -\epsilon_{i,j_2}$ and zeroes elsewhere. The total variation distance between any two rows not involving the $i$-th row remain the same for both $T+E, T-E$. It is enough to check the distance w.r.t $i$-th row is within $\delta$ for all the remaining rows.
\bean
d_{TV}((T\pm E)_i, (T\pm E)_{\hat{i}})&=&\frac{1}{2}\sum_{j=1}^m |T_{i,j}\pm \epsilon_{i,j}-T_{\hat{i},j}|\\
&\le& d_{TV}(T_i, T_{\hat{i}}) + \epsilon^* \\
&\le& d_i + \delta-d_i = \delta.
\eean
Therefore $T\pm E \in \Pd$ for a non-zero perturbation matrix $E$, contradicting the assumption that $T$ is an extreme configuration.
\eprf

\blem\label{lem:tightpoint}
For $k\le3$, every extreme configuration has a tight point.
\elem
\bprf
For $k=2$ case it is clear since the two rows of extreme configuration are tight w.r.t each other, both rows are tight points. For the case when $k=3$, for each $i\in\{1,2,3\}$ define
\[
I_i=\{j\neq i : d_{TV}(T_i,T_j)=\delta\}
\]
as the collection of rows that are tight w.r.t $i$-th row. By Lemma~\ref{lem:deltarow}, each $I_i$ is nonempty, and symmetry of TV distance
implies $j\in I_i$ iff $i\in I_j$. Since
\[
I_1\subseteq\{2,3\},\quad I_2\subseteq\{1,3\},\quad I_3\subseteq\{1,2\}.
\]
W.l.o.g let $2 \in I_1$. This implies that $1 \in I_2$. Since $I_3$ is non-empty, either $1 \in I_3$ or $2 \in I_3$, implying that either row $1$ or $2$ has to be the tight point.

\eprf

\bthm\label{thm:k=2case}
The only possible extreme configurations for the $k=2$ case are the following, up to row and column permutations.
\ben
 \item Single row configuration: 
$\left[\begin{array}{cccc}
     1 & 0 & \cdots & 0 
\end{array}\right].$
\item Two row configurations:
\bean
\label{eq:tworows}
\left[\begin{array}{ccccc}
     1 &  0 & \cdots & 0 \\
     1-\delta & \delta & \cdots & 0
\end{array}\right], \ \left[\begin{array}{ccccc}
     1-\delta & \delta & 0 & \cdots & 0 \\
     1-\delta & 0 & \delta & \cdots & 0
\end{array}\right].
\eean
\een
\ethm
\bprf
 Let us consider the case of extreme configuration with a single row. 
It can be clearly generated as convex combination of $n$ single row configurations unless the matrix is made of singleton row i.e., rows with support in one column. Clearly the singleton row configuration $T$ is an extreme configuration, as the only possible perturbation matrix $E$ that allows for $T \pm E \in \Pd$ is $E=0$.




We will now consider the extreme configuration with two rows and consider two cases (a) with singleton row i.e., a row with support size one and (b) without a singleton row. The singleton row case directly follows from Lemma~\ref{lem:singletonextreme}.

Suppose $T$ has no singleton rows and can be written as
\[
T=
\left[\begin{array}{cccc}
a_1 & a_2 & \cdots & a_m \\
b_1 & b_2 & \cdots & b_m
\end{array}\right].
\]
Assume w.l.o.g that $T$ is localized at column $1$ (due to Lemma~\ref{lem:localisation}), i.e., $
a_i\le\delta, \ b_i\le\delta $ for $i > 1$. 
We first argue that the rows cannot have common support in two or more columns. W.l.o.g suppose $a_1, a_2, b_1, b_2 >0$ then the
perturbation
\[
E=
\left[\begin{array}{cccc}
\epsilon & -\epsilon & 0 & \cdots \\
\epsilon & -\epsilon & 0 & \cdots
\end{array}\right],
\ 
0<\epsilon\le\min\{a_1,a_2,b_1,b_2\},
\]
is feasible from Theorem~\ref{thm:perturb}. Hence an extreme point can have at most one shared support column.  Consequently, $T$ must reduce to the form
\[
T=
\begin{bmatrix}
a_1 & a_2 & \cdots & a_\ell & 0 & \cdots & 0 \\
b_1 & 0 & \cdots & 0 & b_{\ell+1} & \cdots & b_m
\end{bmatrix}.
\]
It is clear to see that $\sum_{i=2}^{\ell}a_i \le \delta$ and $\sum_{i=\ell+1}^{m}b_i \le \delta$ as $d_{TV}(T_1, T_2)\le \delta$ implying that $T \in \Wd$. Therefore by the  decomposition lemma~\ref{lem:decom} for $T$ to be extreme it has to be of the form $E(u_1, u_2)$ for $u_1, u_2 \in \{0\} \cup [2:m]$ as described in the Theorem statement.
\eprf

\bthm\label{thm:k3}
For $k=3$, only possible triple row configurations are of the following form:
\ben
\item With singleton row as shown in Lemma~\ref{lem:singletonextreme} and,
\item Without singleton row:
\bean
\scalebox{0.87}{$\left[\begin{array}{cccccc}
     1-\delta & \delta & 0 & 0 & \cdots & 0 \\
     1-\delta & 0 & \delta & 0 & \cdots & 0 \\
     1-\delta & 0 & 0 & \delta & \cdots & 0
\end{array}\right]$}\\
\scalebox{0.87}{$\left[\begin{array}{ccccc}
     1-2\delta & \delta & \delta&  \cdots & 0 \\
     1-\delta & 0 & \delta & \cdots & 0 \\
     1-\delta & \delta & 0 &  \cdots & 0
\end{array}\right], \left[\begin{array}{cccccc}
     1-2\delta & \delta & \delta & 0 & \cdots & 0 \\
     1-2\delta & 0 & \delta & \delta & \cdots & 0 \\
     1-2\delta & \delta & 0 & \delta & \cdots & 0
\end{array}\right]$}\\
\eean
\een
\ethm

To prove this we introduce a series of Lemmas that reduce the possible supports of extreme configurations to the cases shown in Theorem~\ref{thm:k3}.


\blem \label{lem:supstruct}
Three-row extreme configurations need to have support structure of the following form:
\bea
\label{eq:sup}\left[ \begin{array}{cccccccccc}
     a_1& a_2 & a_3 & a_4 & 0 & 0 & 0 & 0 & \cdots & 0  \\
     b_1 & b_2 & 0 & 0 & b_5 & b_6 & 0 & 0 & \cdots & 0\\
     c_1 & 0 & c_3 & 0 & c_5 & 0 & c_7 & 0 & \cdots & 0\\
\end{array}\right] 
\eea
up to row and column permutation.
\elem
\bprf
Let $T$ be an extreme configuration with three rows and $m$ columns.
Suppose the three rows have a common support in at least two columns, say columns $i,j \geq 1$, choosing $0<\epsilon \leq \min\{a_i,a_j,b_i,b_j,c_i,c_j\}$ and define the perturbation matrix ($E$) by $\epsilon_{1,i}=\epsilon_{2,i}=\epsilon_{3,i}=\epsilon$,
$\epsilon_{1,j}=\epsilon_{2,j}=\epsilon_{3,j}=-\epsilon$,
and $0$ elsewhere. That makes $T\pm E\in\mathcal{P}_\delta$, contradicting extremality. 

Consider two column $i,j \ge 2$ with support in two rows, say $T_2$ and $T_3$, but not in $1$.
Choose $0<\epsilon\le\min\{b_i,c_i,b_j,c_j\}$ and define $\epsilon$ by
$\epsilon_{2,i}=\epsilon_{3,i}=\epsilon$, $\epsilon_{2,j}=\epsilon_{3,j}=-\epsilon$, and $0$ elsewhere.
Then $T\pm E$ remain row--stochastic and satisfy
$d_{\mathrm{TV}}((T\pm E)_2,(T\pm E)_3)\leq\delta$. Now suppose we have two columns $i,j$ with support exactly in row 1, then perturbation $\epsilon_{1,i}=-\epsilon_{1,j}=\epsilon$ with rest of $\epsilon$ values zero is a valid perturbation. Thus, contradicting extremality and forcing the stated support in equation~\eqref{eq:sup}. 
\eprf
\bnote
Arguing as in Lemma~\ref{lem:supstruct}, one can show that for general $k$ the
non-zero columns of an extreme matrix $T \in \Pd$ have pairwise distinct
supports; since $[k]$ has only $2^k-1$ non-empty subsets, it suffices to
consider $m \le 2^k-1$.
\enote
\blem\label{lem:extreme_constr}
Any extreme configuration of the form shown in equation~\eqref{eq:sup} should satisfy at least four distinct constraints from the following list. 
\bean
\{ a_4=0, b_6=0, c_7=0, a_2=b_2,
a_3=c_3, b_5=c_5\\
\label{eq:constr} a_2=b_2=0, a_3=c_3=0, b_5=c_5=0\}.
\eean
\elem
\bprf
Consider a perturbation matrix
\[E=\begin{bmatrix}
     \epsilon_1 & \epsilon_2 & \epsilon_3 & \epsilon_4 & 0 & 0 & 0 & 0 & \cdots & 0  \\
     \epsilon_1 & \epsilon_5 & 0 & 0 & \epsilon_6 & \epsilon_7 & 0 & 0 & \cdots & 0\\
     \epsilon_1 & 0 & \epsilon_8 & 0 & \epsilon_9 & 0 & \epsilon_{10} & 0 & \cdots & 0\\
    
\end{bmatrix}
\]
The rows sums of elements in this matrix need to be zero, this imposes three constraints on the ten variables $\epsilon_1, \cdots, \epsilon_{10}$. At most three more constraints can be imposed by the distance constraints between any pair of rows. Additionally, if fewer than four constraints from the set shown in equation~\eqref{eq:constr} are imposed, it would mean we have 10 variables with fewer than 10 constraints, implying that there exists a non-zero solution for the vector $(\epsilon_1, \cdots, \epsilon_{10})$.  Constraints considered are the only possible options as we are assuming localization in the first column and that columns 2, 3, 5 can have support size of two or zero i.e., it is not possible for $a_2=0$ but $b_2>0$. Also note that the constraint $a_2=b_2$ results in the constraint $\epsilon_2=\epsilon_5$ if the rows 1 and 2 are tight.
\eprf
\bnote
In Lemma~\ref{lem:extreme_constr} the last three equations $a_2=b_2=0$, $a_3=c_3=0$, $b_5=c_5=0$ amount to two constraints each. It is not possible to have $a_2=0$ but $b_2 > 0$ due to  Lemma~\ref{lem:supstruct}. The constraint $a_2=b_2 > 0$ inherently assumes that $d_{TV}(T_1, T_2)=\delta$ as described in the proof of Lemma~\ref{lem:extreme_constr}. Similarly $a_3=c_3$, $b_5=c_5$ assume $d_{TV}(T_1, T_3)=\delta$, $d_{TV}(T_2, T_3)=\delta$ respectively.
\enote
Due to limited space, the proof of Theorem~\ref{thm:k3} is provided in full version of the paper \cite{extreme_ldp_arxiv}. Lemma~\ref{lem:extreme_constr} is used to divide the proof into three broad cases where there are no-singleton column (i.e., $a_4=b_6=c_7=0$), all the singleton $a_4, b_6, c_7 > 0$ and the case where there is atleast one singleton column. 

\section{Some example extreme configurations}

In \cite{extremeLDP}, the authors have defined a collection of matrices that are guaranteed to be $(0, \delta)$-LDP matrices. These matrices are referred to as a star configuration.
\bdefn[Star Configuration]    
 Let $\un{a} = (a_1, \ldots, a_{m-1})$ be a positive tuple of reals such that  $    \sum_{i=1}^{m-1} a_i = 2$. The star configuration corresponding to $\un{a}$ consists of the following points:  
    \begin{itemize}
      \item $T_0 = (a_1 \delta, a_2 \delta, \ldots, a_{m-1} \delta, 1 - 2 \delta)$,
      \item For each $\un{b} \in \{0,1\}^{m-1}$ such that  $
      b_1 a_1 + \cdots + b_{m-1} a_{m-1} = 1$, we have  $
      T_{\un{b}} = (a_1 b_1 \delta, \ldots, a_{m-1} b_{m-1} \delta, 1 - \delta)$.
    \end{itemize}  
\edefn
The point $(a_1 \delta, \ldots, a_{m-1} \delta, 1 - 2 \delta)$ is called the \emph{center} of the configuration. 
We define a key lemma from \cite{extremeLDP} that is used in proving a star configuration is extreme.

    \blem[Lemma~3 \cite{extremeLDP}] \label{lem:starextreme}
    The star configuration for $\un{a}$ is extreme iff the system
    \bean
\sum\limits_{i \in S} \epsilon_i = 0 \text{ for all } S \subseteq [m - 1] \text{ s.t. } \sum\limits_{i \in S} a_i = 1.
\eean
has no non-zero solutions for $\epsilon_1, . . . \epsilon_{m-1}$.
\elem

We use Lemma~\ref{lem:starextreme} to prove few more example star  configurations in Lemmas~\ref{lem:starconfig0},~\ref{lem:starconfig1} and \ref{lem:starconfig2}. 

\blem\label{lem:starconfig0}
Star configuration defined by $
\underbrace{\frac{1}{v_1}, \frac{1}{v_1}, \ldots, \frac{1}{v_1}}_{v_1 \text{ times}},
\quad
\underbrace{\frac{1}{v_2}, \frac{1}{v_2}, \ldots, \frac{1}{v_2}}_{v_2 \text{ times}}
$
such that $\gcd\left(v_1, v_2\right) > 1$ is extreme.
\elem

\bprf
Let $n_1, n_2$ be non-negative integers such that:
\bean
n_1 \frac{1}{v_1} + n_2 \frac{1}{v_2} = 1 \implies n_2 = v_2 \frac{v_1-n_1}{v_1}.
\eean

Let $v_1 = g \hat{v}_1$, $v_2 = g \hat{v}_2$ where $g = \text{gcd}(v_1, v_2)$ and $\text{gcd} (\hat{v}_1, \hat{v}_2) = 1$. 
For $n_2$ to be an integer $n_1$ has to be multiple of $\hat{v}_1$. Let $n_1 = \hat{v}_1$. Since we assume $g > 1$, $n_1, n_2 > 0$ and $n_1 < v_1$ and $n_2<v_2$. Let $S = [1:n_1] \cup [v_1+1:v_1+n_2]$ and $S_{i,j} = S \cup \{ j\} \setminus \{i\}$ for some $i \in [n_1]$ and $j \in [n_1+1:v_1]$.
\bean
\sum\limits_{u \in S} e_u - \sum\limits_{u \in S_{i,j}} e_u = 0 \implies e_i = e_j.
\eean
Therefore $e_1=e_2=\cdots=e_{v_1}=0$. Since $\sum_{i\in [v_1]}{e_i} = 0$. We get $e_i = 0$ for all $i \in [1:v_1]$. Similarly, we can show that $e_i=0$ for all $i \in  [v_1+1:v_1+v_2]$ proving the lemma. 
\eprf

\blem\label{lem:starconfig1}
Star configuration defined by
\[
\underbrace{\frac{1}{v_1}, \frac{1}{v_1}, \ldots, \frac{1}{v_1}}_{x_1 \text{ times}},
\quad
\underbrace{\frac{1}{v_2}, \frac{1}{v_2}, \ldots, \frac{1}{v_2}}_{x_2 \text{ times}}
\quad
\underbrace{\frac{1}{v_3}, \frac{1}{v_3}, \ldots, \frac{1}{v_3}}_{x_3=v_3 \text{ times}}
\]
such that
$\frac{x_1}{v_1}+\frac{x_2}{v_2}=1$ and $\gcd\left(v_3, v_2\right) > 1$, $\gcd\left(v_1, v_3\right) > 1$, $x_1 > \frac{v_1}{\gcd(v_1, v_3)}$ and $x_2 > \frac{v_2}{\gcd(v_2, v_3)}$ is extreme.
\elem
\bprf
Let the index sets for the three groups be $I_1=[1:x_1], I_2=[x_1+1:x_1+x_2], I_3=[x_2+x_1+1:m-1]$ with sizes $x_1, x_2, x_3$ respectively. First, consider the set $I_3$. Since $x_3 = v_3$, the sum of elements in $I_3$ is $v_3 \times \frac{1}{v_3} = 1$. By Lemma~\ref{lem:starextreme}, $\sum_{u \in I_3} e_u = 0$. 
Let $n_1=\frac{v_1}{\text{gcd}(v_1, v_3)}$ and $n_3$ such that $n_1 \frac{1}{v_1} + n_3 \frac{1}{v_3} = 1$.  Clearly $n_1 < x_1$ and $n_3 < v_3$. $S_{1,3}$ be a subset of indices consisting of $n_1$ indices from $I_1$ and $n_3$ indices from $I_3$. Let $S'_{i,j} = S_{1,3} \cup \{j\} \setminus \{i\}$ for some $i \in I_3 \cap S_{1,3}$ and $j \in I_3 \setminus S_{1,3}$ (such $j$ exists because $n_3 < v_3 = x_3$).
\bean\sum\limits_{u \in S_{1,3}} e_u - \sum\limits_{u \in S'_{i,j}} e_u = 0 \implies e_i = e_j.\eean
Since we can swap any elements within $I_3$, all $e_u$ for $u \in I_3$ are equal. Since $\sum_{u \in I_3} e_u = 0$, it follows that $e_u = 0$ for all $u \in I_3$. Now, substituting $e_u = 0$ for $u \in I_3$ into the sum for $S_{1,3}$, we get $\sum_{u \in S_{1,3} \cap I_1} e_u = 0$.
This implies that the sum of any choice of $n_1$ elements from $I_1$ is 0. Using the swapping argument on $I_1$ (letting $S''_{k,l} = (S_{1,3} \cap I_1) \cup \{l\} \setminus \{k\}$ for $k, l \in I_1$), we get $e_k = e_l$ for all $k,l \in I_1$. Since $n_1 e_k = 0$ and $n_1 > 0$, we have $e_u = 0$ for all $u \in I_1$. It can similarly be  shown that $e_u=0$ for all $u\in I_2$ proving the Lemma.
\eprf

\blem\label{lem:starconfig2}
Star configuration defined by
\[ \underline{a}=
\underbrace{\frac{u_1}{v_1}}_{x_1=1},
\quad
\underbrace{\frac{1}{v_1}, \frac{1}{v_1}, \ldots, \frac{1}{v_1}}_{x_2 =v_1-u_1\text{ times}}
\quad
\underbrace{\frac{1}{v_2}, \frac{1}{v_2}, \ldots, \frac{1}{v_2}}_{x_3=v_2 \text{ times}}
\]
such that
$u_1 \le v_1-\frac{v_1}{g}$ where $g = \text{gcd}(v_1, v_2)$, is extreme with the strict inequality required unless $\frac{v_1}{g} = 1$.
\elem
\bprf
Let the index sets be defined as:
$I_1=\{1\}, \quad I_2=\{2,\ldots,v_1-u_1+1\},$ and $I_3=\{v_1-u_1+2,\ldots,v_1-u_1+v_2+1\}.$
By Lemma~\ref{lem:starextreme}, any subset $S$ such that $\sum_{i \in S}a_i = 1$ implies $\sum_{i \in S} e_i = 0$. 
Let $c_1=\frac{v_1}{g}$, and $c_2=v_2 - \frac{v_2}{g}$ then $c_1 \frac{1}{v_1} + c_2 \frac{1}{v_2} = 1$. Since $u_1 \le v_1 - c_1$, $c_1$ entries can be picked from $I_2$ and $c_2$ entries from $I_3$ such that $a_i$s sum to $1$. This results in
\bea
\label{eq:crossconstr}
\sum_{i \in J_2} e_i + \sum_{j \in J_3} e_j = 0
\eea
for any $J_2 \subseteq I_2$ with $|J_2| = c_1$, and $J_3 \subseteq I_3$ with $|J_3| = c_2$. Because $c_2 < |I_3| = v_2$, we can swap any $j \in J_3$ with an element in $I_3 \setminus J_3$ to show that the $e_i=e_j$ for any $i, j \in I_3$. Since we have $\sum_{i \in I_3} a_i = 1$ it follows that $\sum_{i \in I_3} e_i = v_2 e_{x_1+x_2+x_3}=0$ implying $e_j = 0$ for all $j \in I_3$. Therefore the constraint in equation~\eqref{eq:crossconstr} simplifies to:$\sum_{i \in J_2} e_i = 0 \quad \text{for any subset } J_2 \subseteq I_2 \text{ of size } c_1.$

If we have $c_1 < |I_2|$ i.e., $u_1 < v_1-c_1$, we can show that $e_i = 0$ for all $i \in I_2$, similar to how we showed $e_i =0$ for all $i \in I_3$. Since we also have $\sum_{i \in I_1 \cup I_2} e_i = 0$, it forces $e_1=0$ implying that the configuration is extreme whenever $u_1 < v_1-c_1$.

We now consider the case when $c_1=1$ and $u_1=v_1-c_1=v_1-1$. In this case the only subset $J_2$ of size $c_1$ is $I_2$ itself. The equation~\eqref{eq:crossconstr} gives $\sum_{i \in I_2} e_i = 0$, implying $e_2=0$. This further implies that $e_1=0$ as $a_1+a_2=1$ implies $e_1+e_2=0$.
\eprf

\bnote
$v_1=v_2$ in Lemma~\ref{lem:starconfig2} reduces to \cite[Theorem~7]{extremeLDP}.
\enote

\begin{example}
We present example star configurations that result from Lemmas~\ref{lem:starconfig0},~\ref{lem:starconfig1} and \ref{lem:starconfig2} respectively.
\ben
\item $\underline{a} \;=\; \Big(\underbrace{\tfrac{1}{2},\tfrac{1}{2}}_{v_1=2},\;
\underbrace{\tfrac{1}{4},\tfrac{1}{4},\tfrac{1}{4},\tfrac{1}{4}}_{v_2=4}\Big)$ satisfies $\text{gcd}(v_1, v_2)=2 > 1$.
\item $\underline{a} \;=\; \Big(\underbrace{\tfrac{1}{3},\tfrac{1}{3}}_{x_1=2},\;
\underbrace{\tfrac{1}{12},\tfrac{1}{12},\tfrac{1}{12},\tfrac{1}{12}}_{x_2=4},\;
\underbrace{\tfrac{1}{6},\tfrac{1}{6},\tfrac{1}{6},\tfrac{1}{6},\tfrac{1}{6},\tfrac{1}{6}}_{x_3\,=\,v_3\,=\,6}\Big)$ satisfies $\frac{x_1}{v_1}+\frac{x_2}{v_2}=1$, $\text{gcd}(v_1, v_3)=3>1$ and $\text{gcd}(v_2, v_3)=6>1$, $2=x_1>\frac{v_1}{\text{gcd}(v_1, v_3)}=1$, $4=x_2>\frac{v_2}{\text{gcd}(v_2, v_3)}=2$.
\item $\underline{a} \;=\; \Big(\underbrace{\tfrac{2}{6}}_{x_1=1},\;
\underbrace{\tfrac{1}{6},\tfrac{1}{6},\tfrac{1}{6},\tfrac{1}{6}}_{x_2\,=\,v_1-u_1\,=\,4},\;
\underbrace{\tfrac{1}{4},\tfrac{1}{4},\tfrac{1}{4},\tfrac{1}{4}}_{x_3\,=\,v_2\,=\,4}\Big)$ satisfies $2=u_1<v_1-\frac{v_1}{\text{gcd}(v_1, v_3)}=3$.
\end{enumerate}
\end{example}

\paragraph*{AI use statement}
The authors used Claude Opus 5 and Gemini Edu to discuss and verify some of the proofs and results. All mathematical statements, proofs, and conclusions were independently verified by the authors, who take full responsibility for the content of this paper. 

\bibliographystyle{IEEEtran}
\bibliography{LDP.bib}

\begin{appendices}

\section{Proof of Theorem~\ref{thm:k3} \label{sec:theoremk3}}

\blem\label{lem:nosingleton}
Only possible three-row extreme configurations with columns having support sizes two and three are of the form:
\bean
 \scalebox{0.9}{$\begin{bmatrix}
1-2\delta & \delta & \delta & 0 & 0 & \dots & 0 \\
1-2\delta & \delta & 0 & 0 & \delta & \dots & 0 \\
1-2\delta & 0 & \delta & 0 & \delta & \dots & 0
\end{bmatrix}, \begin{bmatrix}
1-2\delta & \delta & \delta & 0 & \dots & 0 \\
1-\delta & \delta & 0 & 0 & \dots & 0 \\
1-\delta & 0 & \delta & 0 & \dots & 0
\end{bmatrix}$}
\eean
up to row and column permutation.
\elem
\bprf
Proof is provided in Appendix~\ref{lem:nosingleton}.
\eprf

\blem \label{lem:allsingleton}
Only possible three-row extreme configuration with all the singleton columns present is of the form:
\bea
\label{eq:allsingleton}
 \scalebox{0.9}{$\left[\begin{array}{cccccc}
     1-\delta & \delta & 0 & 0 & \cdots & 0 \\
     1-\delta & 0 & \delta & 0 & \cdots & 0 \\
     1-\delta & 0 & 0 & \delta & \cdots & 0
\end{array}\right]$} 
\eea
up to row and column permutations.
\elem
\bprf
Proof is provided in Appendix~\ref{sec:lemallsingleton}
\eprf

\blem \label{lem:mixsingleton}
Extreme configuration with three rows without a singleton row are not possible with a mix of singleton columns and columns with support size $\ge 2$.
\elem
\bprf
Proof in provided in Appendix~\ref{sec:lemmixsingleton}
\eprf

\textit{Proof of Theorem~\ref{thm:k3}:}
By Lemma~\ref{lem:supstruct} and Lemma~\ref{lem:extreme_constr}, any extreme configuration with three rows must have a support structure as shown in equation~\eqref{eq:sup} and the variables should satisfy atleast four among the following set of constraints.
\bean
 \{a_4=0, b_6=0, c_7=0, a_2=b_2, a_3=c_3, b_5=c_5, \\ a_2=b_2=0, a_3=c_3=0, b_5=c_5=0\}
\eean
The constraints $a_2=b_2=0$, $a_3=c_3=0$, and $b_5=c_5=0$ effectively set two variables to zero and thus each count as exactly two constraints. We look at four possible cases based on the variables $a_4, b_6,$ and $c_7$. Case 1 where $a_4=b_6=c_7=0$, case 2 where exactly one among $a_4, b_6, c_7$ is non-zero, case 3 where exactly two of them are non-zero and case 4 where $a_4, b_6, c_7 >0$.
\ben
\item Case 1 and 4: Proof for these cases follows from Lemmas~\ref{lem:nosingleton} and \ref{lem:allsingleton} respectively.
\item Cases 2 and 3 follow from Lemma~\ref{lem:mixsingleton}.
\een





\section{Preliminaries}
In this section we present peeling lemma~\ref{lem:peel} that will be used along with the decomposition lemma~\ref{lem:decom} in the proofs of Lemma~\ref{lem:allsingleton},~\ref{lem:nosingleton} and \ref{lem:mixsingleton} in the later sections.




For any $B \subseteq [2:m]$ we define $R_B \in \mathbb{R}^{1 \times m}$ such that $R_B(j) = 0$ for $j \notin B \cup \{1\}$ and  $R_{B}(j)=\delta$ for $j \in B$ and $R_B(1) = 1-|B|\delta$. We will represent $R_{\phi}$ as $R_0$. It follows from the definition that:
\bean
R_0 &=& [1~0~\cdots~0]\\
R_u &=& [1-\delta, 0,~\dots, \delta \text{ (at col } u), \dots, 0],\\
R_{2, 3}&=&[1-2\delta, \delta, \delta, 0, \dots, 0].
\eean
Using these rows, we define the following extreme matrices:
\bean
E(u,v,w)=\begin{bmatrix}
    R_u\\
    R_v\\
    R_w
\end{bmatrix}, 
E(\{u,v\},u,v)=\begin{bmatrix}
    R_{u, v}\\
    R_u\\
    R_v
\end{bmatrix}, \\ E(\{u, v\},\{v, w\},\{u, w\})=\begin{bmatrix}
    R_{u, v}\\
    R_{v, w}\\
    R_{u, w}
\end{bmatrix}.
\eean



\blem[Peeling Lemma]\label{lem:peel}
Let $T \in \Pd$, let $X = E(B_1,B_2,B_3) \in \Pd$ such that $B_i \subseteq [2, m]$ such that $|B_i| \le 2$ and let $\lambda \in (0,1]$ satisfy
\ben
\item $T \ge \lambda X$ entry wise, and
\item $s_i \triangleq \sum_{j=2}^m T_{i,j} \le (1-\lambda)\delta + \lambda |B_i| \delta$ for every $i \in [k]$.
\een
Then it implies that:
\ben
\item[(a)] if $\lambda=1$ for $T$ to be extreme $T=X$,
\item[(b)] if $\lambda \in (0,1)$ then $T' := (T - \lambda X)/(1-\lambda)$ lies in $\Wd$, and $T = \lambda X + (1-\lambda)T'$. In particular, if $T$ is extreme then $T=E(u, v, w)$ for some $u, v, w \in \{0\} \cup [2:m]$
\een
\elem
\bprf
If $\lambda=1$, $T\geq X$ implies that $T-X$ is nonnegative entriywise and every row sums to zero. Thus the corresponding rows of both the matrices are equal. Thus, assume $\lambda\in(0,1)$.
Each row of $T'$ sums to $(1-\lambda)/(1-\lambda) = 1$, and $T' \ge 0$ by (i).
It can be seen that:
\[
  T_{i,1}' = \frac{T_{i,1} - \lambda X_{i,1}}{(1-\lambda)} = \frac{((1-s_i)-\lambda (1-|B_i|\delta))}{1-\lambda} \ge (1-\delta).
\]
Therefore $T' \in \Wd$ and the lemma proof follows from Lemma~\ref{lem:decom}. 
\eprf

\bnote\label{rem:peel}
Note  that is  is enough to check $T\ge\lambda X$ in columns $[2:m]$ and that $s_i \le (1-\lambda)\delta+\lambda |B_i|\delta$. $T_{i1}\ge \lambda X_{i,1}$ follows as shown below:
\bean
T_{i,1}&=&1-s_i \ge 1-(1-\lambda)\delta-\lambda |B_i| \delta \\&=& (1-\lambda)(1-\delta)+\lambda(1-|B_i| \delta) \ge \lambda X_{i,1}.
\eean

\enote

\section{Proof of Lemma~\ref{lem:nosingleton}}

We look at the case of extreme matrices that do not have any singleton columns i.e., we assume $a_4 = b_6 =c_7 = 0$. From Lemma~\ref{lem:extreme_constr} we know that atleast four constraints need to hold among the constraints shown in equation~\ref{eq:constr} should hold. This reduces to the cases of either (a) $a_2=b_2$, (b) $a_3=c_3$ or (c) $b_5=c_5$. Not that for constraint $a_2=b_2$ to force constraints on perturbation matrix, it is also needed that rows 1 and 2 are tight w.r.t each other. Due to the symmetry of these cases we do the proof for $a_2=b_2$ case, the other cases proof follow exactly in similar lines. The structure of the matrix $T$ reduces to the following form, excluding the zero columns up to row/column permutations:
\[
T=\begin{bmatrix}
a_1 & a_2 & a_3 & 0 \\
b_1 & a_2 & 0 & b_5 \\
c_1 & 0 & c_3 & c_5
\end{bmatrix}.
\]




Since $d_{\mathrm{TV}}(T_1, T_2)=\delta$, it implies either $a_3 = \delta \text{ or } b_5 = \delta.$ The matrix form reduces to the following forms for $a_3=\delta$ and $b_5 = \delta$ respectively.
\[
\scalebox{0.9}{$\begin{bmatrix}
1 - a_2 - \delta & a_2 & \delta & 0 \\
1 - a_2 - b_5    & a_2 & 0      & b_5 \\
1 - c_3 - c_5    & 0   & c_3    & c_5
\end{bmatrix}, \ \begin{bmatrix}
1 - a_2 - a_3 & a_2 & a_3 & 0\\
1 - a_2 - \delta   & a_2 & 0      & \delta \\
1 - c_3 - c_5    & 0   & c_3    & c_5
\end{bmatrix}$}
\]
Since the matrices can be obtained by permuting rows 1, 2 and columns 2 and 3, it is enough to look at a single case. We focus therefore on the case that $a_3=\delta$. We will now show that for this matrix either $b_5 \in \{0, \delta\}$ or $c_5 \in \{0, \delta\}$. Otherwise consider perturbation matrix $E$ such that $\epsilon_{2,1}=\epsilon_{3,1}=-\epsilon$, $\epsilon_{2, 4} = \epsilon_{3,4} = \epsilon$ and rest of the entries as zero. It can be seen that $d_{TV}( \hat{T}_2, \hat{T}_3) = d_{TV}(T_2, T_3)$ where $\hat{T} = T \pm E$ and $d_{TV}( \hat{T}_1, \hat{T}_2) = d_{TV}(T_1, T_2)$. $d_{TV}(\hat{T}_1, \hat{T}_3) = \max(a_2+\delta-c_3, c_5 \pm \epsilon) \le \delta$. The case when $c_5=0$ or $b_5=0$ reduces the case of extreme matrices with singleton columns. The case when $c_5=b_5=0$ reduces to extreme matrices with $m=3$. We will now consider two cases (i) $c_5=\delta$ or (ii) $b_5=\delta$.

\begin{enumerate}[labelindent=0pt]
\item $a_3=\delta, c_5=\delta$: In this case, the extreme matrix $T$ has the form \[
T=\begin{bmatrix}
1-a_2-\delta & a_2 & \delta & 0 \\
1-a_2-b_5 & a_2 & 0 & b_5 \\
1-c_3-\delta & 0 & c_3 & \delta
\end{bmatrix}.
\]
Since pairwise distance between any two rows need to be within $\delta$ it follows that $a_2 \le c_3 \le b_5 \le \delta$. The proof now progresses through a series of claims shown below. We will first show that $b_5 = \delta$ or $c_3=b_5$ for the matrix to be extreme. Otherwise $E$ such that $\epsilon_{2,1}=-\epsilon_{2,4}=\epsilon$ is a valid perturbation matrix for $\epsilon \le \min(\delta-b_5, b_5-c_3)$.
\ben
\item $b_5=\delta$ case implies that $c_3=\delta$ or $a_2=c_3$. Otherwise the perturbation matrix $E$ such that $\epsilon_{3,1}=-\epsilon_{3,3}=\epsilon$ results in $T \pm E \in \Pd$ for any $\epsilon \le \min(\delta - c_3, c_3-a_3)$.
\item $c_3=b_5$ implies that $b_5=\delta$ or $b_5=a_2$. Otherwise $E$ such that $\epsilon_{2,1}=\epsilon_{3,1}=-\epsilon_{2,4}=-\epsilon_{3,3}=\epsilon$ results in $T \pm E \in \Pd$ for $\epsilon \le \min(\delta-b_5, b_5-a_2)$.
\een
The possible matrices therefore reduce to the following options:
\bean
\scalebox{0.65}{$
\begin{bmatrix}
1-a_2-\delta & a_2 & \delta & 0 \\
1-a_2-\delta & a_2 & 0 & \delta \\
1-a_2-\delta & 0 & a_2 & \delta
\end{bmatrix}, \begin{bmatrix}
1-a_2-\delta & a_2 & \delta & 0 \\
1-a_2-\delta & a_2 & 0 & \delta \\
1-2\delta & 0 & \delta & \delta
\end{bmatrix}, \begin{bmatrix}
1-a_2-\delta & a_2 & \delta & 0 \\
1-2a_2 & a_2 & 0 & a_2 \\
1-a_2-\delta & 0 & a_2 & \delta
\end{bmatrix}$}.
\eean
In each of these cases, it can be shown that $a_2$ needs to be either $0$ or $\delta$ for $T$ to be extreme. Otherwise, there exists a non-zero perturbation matrix $E$ such that $T \pm E \in \Pd$. $E$ matrix can be formed by placing $\epsilon$'s in the location where $a_2$ is and $-\epsilon$, $-2\epsilon$ in locations that have $1-a_2-\delta$ and $1-2a_2$ respectively.
\item $a_3=\delta, b_5=\delta$: $T$ is of the form:
\[
T=\begin{bmatrix}
1-a_2-\delta & a_2 & \delta & 0 \\
1-a_2-\delta & a_2 & 0 & \delta \\
1-c_3-c_5 & 0 & c_3 & c_5
\end{bmatrix}.
\]
$d_{TV}(T_1, T_3) = \max(a_2+\delta-c_3, c_5)$ and $d_{TV}(T_2, T_3)=\max(a_2+\delta-c_5, c_3)$ and $a_2 \le \min(c_3, c_5)$. For a tight pair to exist for row 3 (from Lemma~\ref{lem:deltarow}) if $a_2+\delta \le c_3+c_5$ either $c_5=\delta$ or $c_3=\delta$ and for $a_2+\delta>c_3+c_5$ need either $a_2=c_3$ or $a_2=c_5$.
\ben
\item $a_2+\delta \le c_3+c_5$, $c_5=\delta$ implies $c_3=\delta$ or $a_2=c_3$. Otherwise perturbation with $\epsilon_{3,1}=-\epsilon_{3,3} = \epsilon$ such that $\epsilon \le \min(\delta-c_3, c_3-a_2)$ results in $T \pm E \in \mathcal{P}$.
\item  $a_2+\delta \le c_3+c_5$, $c_3=\delta$ implies $c_5=\delta$ or $c_5=a_2$. The proof is similar to earlier case as the matrix in this case is permutation of the matrix seen in earlier case.
\item $a_2+\delta > c_3+c_5$, $a_2=c_3$ implies $c_5=\delta$ or $c_5=a_2$. Otherwise perturbation with $\epsilon_{3,1}=-\epsilon_{3,4} = \epsilon$ such that $\epsilon \le \min(\delta-c_5, c_5-a_2)$ results in $T \pm E \in \mathcal{P}$.
\item $a_2+\delta > c_3+c_5$, $a_2=c_5$ implies $c_3=\delta$ or $c_3=a_2$. Prove follows in same lines as earlier case.
\een
The possible matrices therefore reduce to the following options:
\bean
\scalebox{0.65}{$
\begin{bmatrix}
1-a_2-\delta & a_2 & \delta & 0 \\
1-a_2-\delta & a_2 & 0 & \delta \\
1-a_2-\delta & 0 & a_2 & \delta
\end{bmatrix}, \begin{bmatrix}
1-a_2-\delta & a_2 & \delta & 0 \\
1-a_2-\delta & a_2 & 0 & \delta \\
1-2\delta & 0 & \delta & \delta
\end{bmatrix}, \begin{bmatrix}
1-a_2-\delta & a_2 & \delta & 0 \\
1-2a_2 & a_2 & 0 & a_2 \\
1-a_2-\delta & 0 & a_2 & \delta
\end{bmatrix}$}.
\eean
As shown earlier, these matrices can be extreme only if $a_2=0$ or $a_2=\delta$.
\end{enumerate}

\section{ Proof of Lemma~\ref{lem:allsingleton} \label{sec:lemallsingleton}}
In this case we look at extreme matrices $T$ (see equation~\eqref{eq:sup}) such that $a_4 > 0$, $b_6 > 0$ and $c_7 > 0$. From Lemma~\ref{lem:extreme_constr} we know that at least four constraints out of the following constraints need to hold:
\bean
\scalebox{0.9}{$
 \{a_2=b_2, a_3=c_3, b_5=c_5, a_2=b_2=0, a_3=c_3=0, b_5=c_5=0\}$.}
\eean
This can be divided into the following two cases:
\ben
\item Exactly two out of the following three conditions hold $\{a_2=b_2=0, a_3=c_3=0, b_5=c_5=0\}$. For example, $a_2=b_2=0$ and $a_3=c_3=0$ imply four constraints. Due to symmetry it is enough to look at the example case alone.
\item One of the three conditions $\{a_2=b_2=0, a_3=c_3=0, b_5=c_5=0\}$ hold and two out to the three conditions hold: $\{a_2=b_2, a_3=c_3, b_5=c_5\}$. For example, $a_2=b_2=0$, $a_3=b_3$ and $b_5=c_5$ result in four constraints and it is enough to look at this case alone due to symmetry.
\een
We proceed with the first case followed by the second one.
\ben
\item Case 1: $a_2=b_2=0, a_3 = c_3 = 0$. In this case, the matrix $T$ will be of the form:
\bean
T =\begin{bmatrix}
1-a_4 & 0 & 0 & a_4 & 0 & 0 & 0\\
1-b_5-b_6 & 0 &  0& 0 & b_5 & b_6 & 0 \\
1-c_5-c_7 & 0 & 0  & 0 & c_5 & 0 & c_7 
\end{bmatrix}.
\eean 
From our assumption, $a_4, b_6, c_7 > 0$. Since $T \in \Pd$ it forces $b_5+b_6 \le \delta$, $c_5+c_7 \le \delta$, $a_4 \le \delta$. Therefore $T \in \Wd$ and from Lemma~\ref{lem:decom} it follows that  for $T$ to be extreme, it needs to be of the form $E(u, v, w)$ for $u, v, w \in \{0\} \cup [2:m]$.
\item Case 2: $a_2=b_2=0, a_3 = c_3, b_5 = c_5$. In this case $T$ is of the form:
\bean
T =\begin{bmatrix}
1-a_3-a_4 & 0 & a_3 & a_4 & 0 & 0 & 0\\
1-b_5-b_6 & 0 &  0& 0 & b_5 & b_6 & 0 \\
1-a_3-b_5-c_7 & 0 & a_3  & 0 & b_5 & 0 & c_7 
\end{bmatrix}.
\eean 
Since we also have that $d_{TV}(T_1, T_3)=d_{TV}(T_2, T_3) = \delta$, it follows that $\max(a_4, b_5+c_7)=\max(b_6, a_3+c_7)=\delta$. If $s_3=a_3+b_5+c_7 \le \delta$ then $T \in \Wd$ and the Lemma proof follows from Lemma~\ref{lem:decom}.  Consider the case where $s_3 > \delta$. It can also be seen that $s_3 \le 2\delta$ as $a_3 \le \delta$ and $b_5+c_7 \le \delta$.

We will now show that $T$ along with $X=E(3, 5, \{3,5\})$ satisfies the necessary conditions required for the Lemma~\ref{lem:peel} to go through for $\lambda = \frac{s_3-\delta}{\delta}$ where 
\bean
\lambda X =\begin{bmatrix}
    \lambda(1-\delta)& 0 & s_3-\delta & 0 & 0 & 0 & 0\\
     \lambda(1-\delta)& 0 & 0 & 0 & s_3-\delta   & 0 & 0\\
      \lambda(1-2\delta)& 0 & s_3-\delta & 0 & s_3-\delta   & 0 & 0\\
\end{bmatrix}
\eean
It can be seen that $a_3\ge s_3-\delta$ as $b_5+c_7 \ delta$ and $b_5\ge s_3-\delta$ as $a_3+c_7 \le\delta$. 
\bean
s_1 = a_3+a_4 \le (1-\lambda) \delta+\lambda \delta = \delta\\
s_2=b+5+b_6\le (1-\lambda) \delta+\lambda \delta = \delta\\
s_3 = a_3+b_5+c_7 \le (1-\lambda) \delta+2\lambda \delta=s_3.
\eean
 This also ensures that $T_{i,1} \ge \lambda X_{i,1}$ by remark~\ref{rem:peel}. From Lemma~\ref{lem:peel} it follows that $T=X$ or $T=E(u, v, w)$ for some $u, v, w \in \{0\} \cup [2:m]$. Since we assume that $a_4, b_6, c_7  > 0$, $T \ne X$. It has to be of the form $T=E(u, v, w)$ for some $u, v, w \in \{0\} \cup [2:m]$.
\een

\section{Proof of Lemma~\ref{lem:mixsingleton} \label{sec:lemmixsingleton}}

We will first consider the case where exactly two of $\{a_4, b_6, c_7\}$ are positive, and the third is zero, and later consider the case where exactly one of them is positive and the rest zero.

\subsection{Two Singleton Column}
Without loss of generality, we assume $a_4>0, b_6>0$ and $c_7=0$. From Lemma~\ref{lem:extreme_constr} we know that at least three out the following constraints need to hold.
\bean
\scalebox{0.9}{$
 \{a_2=b_2, a_3=c_3, b_5=c_5, a_2=b_2=0, a_3=c_3=0, b_5=c_5=0\}$.}
\eean
This results in the following seven cases. 
\ben
\item $a_2=b_2=0, a_3=c_3$,
\item $a_2=b_2=0, b_5=c_5$, 
\item $a_3=c_3=0, a_2=b_2$,
\item $a_3=c_3=0, c_5=b_5$,
\item $b_5=c_5=0, a_2=b_2$, 
\item $b_5=c_5=0, a_3=c_3$, and 
\item $a_2=b_2, a_3=c_3, b_5=c_5$.
\een

\ben
\item Case 1 and 2: These two cases are the same, as the $T$ matrix in one case can be obtained by row/column permuting that of the other. We will therefore look at the scenario when $a_2=b_2=0$ and $a_3=c_3$. The matrix $T$ is of the form:
\bean
T =\begin{bmatrix}
1-a_4-a_3 & 0 & a_3 & a_4 & 0 & 0 & 0\\
1-b_5-b_6 & 0 &  0& 0 & b_5 & b_6 & 0 \\
1-a_3-c_5 & 0 & a_3  & 0 & c_5 & 0 & 0 
\end{bmatrix}.
\eean 
It is clear to see that $a_3+a_4\le \delta$, $b_5+b_6\le\delta$. 
\ben
\item If $a_3 + c_5 \le \delta$ then $T \in \Wd$ and by Lemma~\ref{lem:decom} it follows that $T = E(u, v, w)$ for some $u, v, w \in \{0\} \cup [2:m]$. Since we know only two-singleton column are present in $T$ it  forces $T$ to have a singleton row. 
\item Consider the case when $s_3 = a_3+c_5 > \delta$. It is clear to see that $s_3 \le 2 \delta$. We will use Lemma~\ref{lem:peel} to prove that $T$ cannot be extreme matrix with exactly two singleton column and no-singleton row.  Let $X=E(3, 5, \{3, 5\})$ and $\lambda = \frac{s_3-\delta}{\delta}$ then
\bean
\lambda X = \begin{bmatrix}
\lambda(1-\delta) & 0 & s_3-\delta & 0 & 0 & 0 & 0\\
\lambda(1-\delta)  & 0 &  0& 0 & s_3-\delta & 0 & 0 \\
\lambda(1-2\delta)  & 0 & s_3-\delta  & 0 & s_3-\delta & 0 & 0 
\end{bmatrix}.
\eean
It is clear to see that the entries in $T$ are greater than the entries seen in $\lambda X$ at columns $[2:7]$. We will now show that the other requirements of the Lemma are satisfied i.e., 
\bean
s_1 = a_3+a_4 \le (1-\lambda) \delta + \lambda \delta = \delta,\\
s_2 = b_5+b_6 \le (1-\lambda) \delta + \lambda \delta \le \delta,\\
s_3 = a_3+c_5 = (1-\lambda) \delta + \lambda 2 \delta = s _3.
\eean
   This also ensures that $T_{i,1} \ge \lambda X_{i,1}$ by remark~\ref{rem:peel}. Therefore by applying Lemma~\ref{lem:peel} it follows that either $T=X$ when $s_3=2\delta$ or $T = E(u, v, w)$ for $u, v, w \in \{0\} \cup [2:m]$. $T$ can't be equal to $X$ in this case as we assume $a_4, b_6 >0$. If $T = E(u, v, w)$ it forces one of the rows  be a singleton row.
\een
\item Case 3 and 5: Since case 3 and 5 result in matrices with same support upto row-column permutation, we focus on case 3 where $a_3=c_3=0$ and $a_2=b_2$. Since row 1 and 2 are tight w.r.t each other we have $\max(a_4, b_5+b_6)=\delta$. So either $a_4=\delta$ or $b_5+b_6=\delta$.
\bean
T =\begin{bmatrix}
1-a_4-a_2 & a_2 & 0 & a_4 & 0 & 0 & 0\\
1-b_5-b_6-a_2 & a_2 &  0& 0 & b_5 & b_6 & 0 \\
1-c_5 & 0 & 0 & 0 & c_5 & 0 & 0 
\end{bmatrix}.
\eean 
It is clear to see that $a_2+a_4\le \delta$, $c_5 \le \delta$ and let $s_2 = a_2+b_5+b_6$. 
\ben
\item If $s_2 \le \delta$ it follows that $T \in \Wd$ and from Lemma~\ref{lem:decom} $T=E(u, v, w)$ f or $u, v, w \in \{0\} \cup [2:m]$
\item If $s_2 > \delta$  we will show that the peeling lemma can be applied with  $X=E(2,\{2,5\},5)$ and $\lambda = \frac{s_2-\delta}{\delta}$.
\bean
\lambda X = \begin{bmatrix}
\lambda(1-\delta) &  s_2-\delta & 0 & 0 & 0 & 0 & 0\\
\lambda(1-2\delta)  & s_2-\delta &  0& 0 & s_2-\delta & 0 & 0 \\
\lambda(1-\delta)  & 0 &0  & 0 & s_2-\delta & 0 & 0 
\end{bmatrix}.
\eean
It follows that $a_2 \ge s_2-\delta$ as  $b_5+b_6 \le \delta$ and $c_5 \ge s_2-\delta$ as $d_{TV}(T_2, T_3)\le \delta$. We will now show that the other requirements of the Lemma are satisfied i.e., 
\bean
s_1 = a_2+a_4 \le (1-\lambda) \delta + \lambda \delta = \delta,\\
s_2 = a_2+b_5+b_6 \le (1-\lambda) \delta + 2\lambda \delta =s_2,\\
s_3 = c_5 = (1-\lambda) \delta + \lambda \delta.
\eean
This also ensures that $T_{i,1} \ge \lambda X_{i,1}$ by remark~\ref{rem:peel}. Therefore by applying Lemma~\ref{lem:peel} it follows that either $T=X$ when $s_2=2\delta$ or $T = E(u, v, w)$ for $u, v, w \in \{0\} \cup [2:m]$. $T$ can't be equal to $X$ in this case as we assume $a_4, b_6 >0$. If $T = E(u, v, w)$ it forces one of the rows  be a singleton row. 
\een

\item Case 4 and 6: Since both these cases result in matrices with the same support up to row-column permutation. We focus on case 6 where $b_5=c_5=0$, $a_3=c_3$. The matrix $T$ is of the form:
\[
\begin{bmatrix}
1-a_2-a_3-a_4&  a_2&  a_3 & a_4 & 0 \\
1-b_2-b_6 &   b_2& 0 & 0 & 0 & b_6 & 0  \\
1-a_3& 0 & a_3&0 & 0 & 0 & 0 
\end{bmatrix}
\]
It is clear to see that $b_2+b_6\le \delta$, $a_3 \le \delta$ and let $s_1 = a_2+a_3+a_4$. 
\ben
\item If $s_1 \le \delta$ it follows that $T \in \Wd$ and from Lemma~\ref{lem:decom} $T=E(u, v, w)$ f or $u, v, w \in \{0\} \cup [2:m]$
\item If $s_1 > \delta$  we will show that the peeling lemma can be applied with  $X=E(\{2, 3\},2,3)$ and $\lambda = \frac{s_1-\delta}{\delta}$.
\bean
\lambda X = \begin{bmatrix}
\lambda(1-2\delta) &  s_1-\delta & s_1-\delta & 0 & 0 & 0 & 0\\
\lambda(1-\delta)  & s_1-\delta &  0& 0 & 0 & 0 & 0 \\
\lambda(1-\delta)  & 0 & s_1-\delta  & 0 & 0 & 0 & 0 
\end{bmatrix}.
\eean
It follows that $a_2 \ge s_1-\delta$ as  $a_3+a_4 \le \delta$ and $b_2 \ge s_1-\delta$ as $d_{TV}(T_1, T_2)\le \delta$. We will now show that the other requirements of the Lemma are satisfied i.e., 
\bean
s_1 = a_2+a_3+a_4 \le (1-\lambda) \delta + 2\lambda \delta = s_1,\\
s_2 = b_2+b_6 \le (1-\lambda) \delta + \lambda \delta =\delta,\\
s_3 = a_3 = (1-\lambda) \delta + \lambda \delta =\delta.
\eean
This also ensures that $T_{i,1} \ge \lambda X_{i,1}$ by remark~\ref{rem:peel}. Therefore by applying Lemma~\ref{lem:peel} it follows that either $T=X$ when $s_1=2\delta$ or $T = E(u, v, w)$ for $u, v, w \in \{0\} \cup [2:m]$. $T$ can't be equal to $X$ in this case as we assume $a_4, b_6 >0$. If $T = E(u, v, w)$ it forces one of the rows  be a singleton row. 
\een

\item Case 7: $a_2 = b_2, a_3 = c_3, b_5 = c_5$. $T$ in this case is of the form:
\bean
T =\begin{bmatrix}
1-a_4-a_2-a_3 & a_2 & a_3 & a_4 & 0 & 0 & 0\\
1-a_2-b_5-b_6 & a_2 &  0& 0 & b_5 & b_6 & 0 \\
1-a_3-b_5 & 0 & a_3  & 0 & b_5 & 0 & 0 
\end{bmatrix}.
\eean 
Since every pairs of rows are tight w.r.t each other, we conclude $\max\{b_5+b_6,a_3+a_4\} = \max\{a_3,a_2+b_6\}=\max\{a_2+a_4,b_5\} =\delta$. As $a_4, b_6 > 0$ it follows that $a_3, b_5 < \delta$. Therefore the tightness conditions reduce to: $a_2+b_6=a_2+a_4=\delta$ and $\max(b_5+b_6, a_3+a_4)=\delta$.
\ben
\item $b_5+b_6=\delta$: $T$ is of the form shown below:
\bean
\begin{bmatrix}
1-\delta-a_3 & a_2 & a_3 & \delta-a_2 & 0 & 0 & 0\\
1-a_2-\delta & a_2 &  0& 0 & a_2 & \delta-a_2 & 0 \\
1-a_3-a_2 & 0 & a_3  & 0 & a_2 & 0 & 0 
\end{bmatrix}.
\eean 
For $T$ to be extreme it is required that $a_3=a_2$ or $a_3=0$ otherwise the perturbation matrix $E$ such that $\epsilon_{1,1}=-\epsilon_{1,3}=\epsilon_{3,1}=-\epsilon_{3,3}$ results in $T \pm E \in \Pd$. The possible matrices reduces to following two options:
\bean
\begin{bmatrix}
1-\delta-a_2 & a_2 & a_2 & \delta-a_2 & 0 & 0 & 0\\
1-a_2-\delta & a_2 &  0& 0 & a_2 & \delta-a_2 & 0 \\
1-2a_2 & 0 & a_2  & 0 & a_2 & 0 & 0 
\end{bmatrix},\\ \begin{bmatrix}
1-\delta& a_2 &  & \delta-a_2 & 0 & 0 & 0\\
1-a_2-\delta & a_2 &  & 0 & a_2 & \delta-a_2 & 0 \\
1-a_2 & 0 &   & 0 & a_2 & 0 & 0 
\end{bmatrix}.
\eean
Since $a_4, b_6 >0$, we have $a_2 < \delta$. If $a_2>0$ there is a perturbation matrix $E$ defined by setting $\epsilon$ where $a_2$ is present, $-\epsilon$ at locations where $\delta-a_2$ or $1-a_2-\delta$ is present and $-2\epsilon$ at locations where $1-2a_2$ is present that results in $T \pm E \in \Pd$.
\item $a_3+a_4=\delta$. $T$ is of the form:
\bean
\begin{bmatrix}
1-\delta-a_2 & a_2 & a_2 & \delta-a_2 & 0 & 0 & 0\\
1-b_5-\delta & a_2 &  0& 0 & b_5 & \delta-a_2 & 0 \\
1-a_2-b_5 & 0 & a_2  & 0 & b_5 & 0 & 0 
\end{bmatrix}.
\eean
For $T$ to be an extreme matrix either $b_5=0$ or $b_5=a_2$. Otherwise the perturbation matrix $E$ such that $\epsilon_{2,1}=\epsilon_{3,1}=-\epsilon_{2,5}=-\epsilon_{3,5}=\epsilon$ results in $T \pm E \in \Pd$. The possible forms of $T$ matrices has entries that are functions of $a_2$ and unless $a_2=0$, there exists a perturbation matrix $E$ such that $T \pm E \in \Pd$ similar to the earlier case.
\een
\een

\subsection{One singleton column}
We will now consider the case where exactly two of $\{a_4, b_6, c_7\}$ are zero, and the other is non-zero. Without loss of generality we assume $a_4=0, b_6=0$ and $c_7>0$. From Lemma~\ref{lem:extreme_constr} we know that atleast two out the following constraints need to hold.
\bean
\scalebox{0.9}{$
 \{a_2=b_2, a_3=c_3, b_5=c_5, a_2=b_2=0, a_3=c_3=0, b_5=c_5=0\}$.}
\eean
This results in the following cases. 
\ben
\item $a_2=b_2=0$, 
\item $a_3=c_3=0$,
\item $b_5=c_5=0$,
\item $a_3=c_3, c_5=b_5$,
\item $b_5=c_5, a_2=b_2$, and 
\item $a_2=b_2, a_3=c_3$.
\een

Cases 2 and 3 are symmetric to each other up to row/column permutation. 
Similarly, cases 5 and 6 are similar to each other. So it is enough to look at cases 1, 2, 4 and 6.
\subsubsection{Case 1:  $a_2=b_2=0$ }
 $T$ is of the form:
\begin{equation*}
T = \begin{bmatrix}
1-a_3 & 0 & a_3 & 0 & 0 & 0 & 0 \\
1-b_5 & 0 & 0 & 0 & b_5 & 0 & 0 \\
1-c_3-c_5-c_7 & 0 & c_3 & 0 & c_5 & 0 & c_7
\end{bmatrix}
\end{equation*}
Let $s_3=c_3+c_5+c_7$. If $s_3 \le \delta$ then $T \in \Wd$ and Lemma~\ref{lem:decom} can be applied to show that $T$ is of the form $E(u, v, w)$ for $u, v, w \in \{0\}. \cup [2:m]$. For the case when $s_3 > \delta$, we will show that Lemma~\ref{lem:peel} can be applied with  $X=E(3, 5, \{3,5\}$ and $\lambda = \frac{s_3-\delta}{\delta}$. \bean
\lambda X = \begin{bmatrix}
\lambda(1-\delta) &  0 & s_3-\delta & 0 & 0 & 0 & 0\\
\lambda(1-\delta)  & 0 &  0& 0 & s_3-\delta & 0 & 0 \\
\lambda(1-2\delta)  & 0 & s_3-\delta  & 0 & s_3-\delta & 0 & 0 
\end{bmatrix}.
\eean
It follows that $a_3 \ge s_3-\delta$ as $d_{TV}(T_1, T_3)\le \delta$, $c_3 \ge s_3 - \delta$ as $c_5+c_7 \le \delta$, $b_5 \ge s_3-\delta$ as $d_{TV}(T_3, T_2)\le \delta$ and $c_5 \ge s_3-\delta$ as $c_3+c_7 \le \delta$. We will now show that the other requirements of the Lemma are satisfied i.e., 
\bean
s_1 = a_3 \le (1-\lambda) \delta + \lambda \delta = \delta\\
s_2 = b_5 \le (1-\lambda) \delta + \lambda \delta =\delta,\\
s_3 = c_3+c_5+c_7 = (1-\lambda) \delta + 2\lambda \delta =s_3.
\eean
This also ensures that $T_{i,1} \ge \lambda X_{i,1}$ by remark~\ref{rem:peel}. Therefore by applying Lemma~\ref{lem:peel} it follows that either $T=X$ when $s_3=2\delta$ or $T = E(u, v, w)$ for $u, v, w \in \{0\} \cup [2:m]$. $T$ can't be equal to $X$ in this case as we assume $c_7 >0$. If $T = E(u, v, w)$ it forces two of the rows  to be singleton row. 

\subsubsection{Case 2:  $a_3=b_3=0$} $T$ is of the form:
\begin{equation*}
T = \begin{bmatrix}
1-a_2 & a_2 & 0 & 0 & 0 & 0 & 0 \\
1-b_2-b_5 & b_2 & 0 & 0 & b_5 & 0 & 0 \\
1-c_5-c_7 & 0 & 0 & 0 & c_5 & 0 & c_7
\end{bmatrix}
\end{equation*}
Since $T \in \Pd$ it follows that $a_2 \le \delta$,  $c_5+c_7 \le \delta$.Let $s_2=b_2+b_5$. If $s_2 \le \delta$ then $T \in \Wd$ as $a_2 \le \delta$,  $c_5+c_7\le \delta$ and Lemma~\ref{lem:decom} can be applied to show that $T$ is of the form $E(u, v, w)$ for $u, v, w \in \{0\}. \cup [2:m]$. For the case when $s_2 > \delta$, we will show that Lemma~\ref{lem:peel} can be applied with  $X=E(2, \{2,5\}, 5)$ and $\lambda = \frac{s_2-\delta}{\delta}$. 
\bean
\lambda X = \begin{bmatrix}
\lambda(1-\delta) & s_2-\delta & 0 & 0 & 0 & 0 & 0\\
\lambda(1-2\delta)  & s_2-\delta &  0& 0 & s_2-\delta & 0 & 0 \\
\lambda(1-\delta)  & 0 & 0 & 0 & s_2-\delta & 0 & 0 
\end{bmatrix}.
\eean
It follows that $a_2 \ge s_2-\delta$ as $d_{TV}(T_1, T_2)\le \delta$, $b_2 \ge s_2-\delta$ as $b_5 \le \delta$, $b_5 \ge s_2-\delta$ as $b_3 \le \delta$  and $c_5 \ge s_3 - \delta$ as $d_{TV}(T_3, T_2)\le \delta$. We will now show that the other requirements of the Lemma~\ref{lem:peel} are satisfied i.e., 
\bean
s_1 = a_2 \le (1-\lambda) \delta + \lambda \delta = \delta\\
s_2 = b_2+b_5 \le (1-\lambda) \delta + 2\lambda \delta =s_2,\\
s_3 = c_5+c_7 \le (1-\lambda) \delta + \lambda \delta =\delta.
\eean
This also ensures that $T_{i,1} \ge \lambda X_{i,1}$ by Remark~\ref{rem:peel}. Therefore by applying Lemma~\ref{lem:peel} it follows that either $T=X$ when $s_2=2\delta$ or $T = E(u, v, w)$ for $u, v, w \in \{0\} \cup [2:m]$. $T$ can't be equal to $X$ in this case as we assume $c_7 >0$. If $T=E(u,v,w)$, then its three rows cannot be distinct.
Indeed, if $u,v,w$ were distinct, at least two would be nonzero,
giving at least two singleton columns. This contradicts the
assumption that $T$ has exactly one singleton column.

\subsubsection{Case 4:  $b_5 = c_5,a_3=c_3$ }

\[
\begin{bmatrix}
1-a_2-a_3 & a_2 & a_3 & 0 & 0 & 0 & 0\\
1-b_2-b_5 & b_2 &  0& 0 & b_5 & 0 & 0 \\
1-a_3-b_5-c_7 & 0 & a_3  & 0 & b_5 & 0 & c_7 
\end{bmatrix}
\]
Since $c_7 > 0$, it follows that $a_3 < \delta$ and $b_5 < \delta$. Since, rows 1-3 and 2-3 are tight, we obtain 2 conditions: max$\{a_2,b_5+c_7\}=\delta$ and max$\{a_3+c_7,b_2\}=\delta$. This results in four cases. We will show that the only possible options for $T$ to be extreme are that $c_7=\delta, a_3=b_5=0$ and $a_2, b_2 \in \{0, \delta\}$.
\ben
\item $a_2 = b_2 = \delta$

We claim that either $a_3 = 0$ or $a_3+c_7=\delta$. Otherwise, then the perturbation matrix $E$:
\[
\begin{bmatrix}
-\epsilon & 0 & \epsilon & 0 & 0 & 0 & 0\\
0 & 0 &  0& 0 & 0 & 0 & 0 \\
-\epsilon & 0 & \epsilon  & 0 & 0 & 0 & 0
\end{bmatrix}
\]
results in $T \pm E \in \Pd$.
\ben
\item When $a_3=0$, the matrix $T$ reduces to the form:
\[
\begin{bmatrix}
1-\delta & \delta & 0 & 0 & 0 & 0 & 0\\
1- \delta-b_5 & \delta &  0& 0 & b_5 & 0 & 0 \\
1-b_5-c_7 & 0 & 0 & 0 & b_5 & 0 & c_7
\end{bmatrix}
\]
We can show that $b_5=0$ or $b_5+c_7=\delta$. Otherwise, perturbation matrix $E$  such that $\epsilon_{2,1}=\epsilon_{3,1}=-\epsilon_{2,5}=-\epsilon_{3,5}=\epsilon$ results in $T\pm E \in \Pd$. For the case when $b_5=0$ through perturbation matrix $E$ such that $\epsilon_{3,1}=-\epsilon_{3,7}=\epsilon$ it can be argued that $c_7=\delta$. For the case when $b_5+c_7=\delta$, entries of matrix $T$ are functions of $c_7$. With perturbation matrix $E$ defined such that $\epsilon_{2,1}=-\epsilon_{2,5}=-\epsilon_{3,5}=\epsilon_{3,7}=\epsilon$, we get that $c_7=\delta$. In this case the only  possible options for $T$ matrix are $a_2=b_2=\delta, a_3=b_5=0, c_7=\delta$.
\item When $a_3+c_7=\delta$, the matrix $T$ reduces to the form:
\begin{align}
\begin{bmatrix}
1-2\delta+c_7 & \delta & \delta-c_7 & 0 & 0 & 0 & 0\\
1- \delta-b_5 & \delta &  & 0 & b_5 & 0 & 0 \\
1-\delta-b_5 & 0 & \delta-c_7 & 0 & b_5 & 0 & c_7
\end{bmatrix}.
\end{align}
We can show that either $b_5=0$ or $b_5+c_7=\delta$, as perturbation $E$ such that $\epsilon_{2,1}=\epsilon_{3,1}=-\epsilon_{2,5}=-\epsilon_{2,5}=\epsilon$ results in $T\pm E \in \Pd$. For the case when $b_5=0$, the perturbation matrix $E$ defined by $\epsilon_{1,1}=-\epsilon_{1,3}=-\epsilon_{3,3}=\epsilon_{3,7}$ results in $T\pm E\in \Pd$ unless $c_7=\delta$. Similarly, for the case when $b_5 = \delta-c_7$, the perturbation matrix defined by 
\[
E=\begin{bmatrix}
\epsilon & 0 & -\epsilon & 0 & 0 & 0 & 0\\
\epsilon & 0 &  & 0 & -\epsilon & 0 & 0 \\
\epsilon & 0 & -\epsilon & 0 & -\epsilon & 0 & \epsilon
\end{bmatrix}
\]
results in $T \pm E \in \Pd$ unless $c_7 = \delta$.
\een
\item  The cases (a) $a_2=\delta, a_3+c_7=\delta$ and (b) $b_2=\delta, b_5+c_7=\delta$ are the same. So we focus on the first scenario. In this case $T$ is of the form:
\[
T=\begin{bmatrix}
1-2\delta+c_7 & \delta & \delta-c_7 & 0 & 0 & 0 & 0\\
1-b_2-b_5 & b_2 &  0& 0 & b_5 & 0 & 0 \\
1-\delta-b_5 & 0 & \delta-c_7  & 0 & b_5 & 0 & c_7 
\end{bmatrix}
\]
Similar to the earlier case it can be shown that $b_5=0$ or $b_5+c_7=\delta$ otherwise the perturbation matrix $E$ such that $\epsilon_{2,1}=\epsilon_{3,1}=-\epsilon_{2,5}=-\epsilon_{3,5}=\epsilon$ is such that $T \pm E \in \Pd$. For the case when $b_5=0$ it can be shown that $b_2=0$ or $b_2+c_7=\delta$ otherwise $E$ such that $\epsilon_{2,1}=-\epsilon_{2,2}$ results in $T \pm E \in \Pd$. For both $b_2=0$ and $b_2+c_7=\delta$ cases, we can show that for $T$ to be extreme it is required that $c_7=\delta$. Further it can be shown that $c_7=\delta$ by defining perturbation matrix such that it has $\epsilon$ in place of $c_7, 1-2\delta+c_7$ and $-\epsilon$ in place of $\delta-c_7$. 

\item $b_5+c_7=\delta, a_3+c_7=\delta$. In this case the matrix $T$ is in the form shown below:
\[
\begin{bmatrix}
1-a_2-\delta+c_7 & a_2 & \delta-c_7 & 0 & 0 & 0 & 0\\
1-b_2-\delta+c_7 & b_2 &  0& 0 & \delta-c_7 & 0 & 0 \\
1-2\delta+c_7 & 0 & \delta-c_7  & 0 & \delta-c_7 & 0 & c_7 
\end{bmatrix}
\]
W.l.o.g we can assume that $a_2 \ge b_2$ as the other case can be thought of as a row/column permutation of this case. For $T$ to be extreme, it is needed that $b_2 \in  \{0, \delta\}$ or $a_2 \in \{0,\delta\}$. Otherwise perturbation matrix $E$ such that $\epsilon_{1,1}=\epsilon_{2,1}=-\epsilon_{1,2}=-\epsilon_{2,2}=\epsilon$ results in $T \pm E \in \Pd$. $b_2=0$ forces $a_2=0$ and $a_2=0$ forces $b_2=0$ as we assume only one singleton column. $b_2=\delta$ forces $a_2=\delta$ as $a_2 \ge b_2$ and $a_2 \le \delta$. When $a_2=\delta$ it can be seen that  $b_2+c_7 \ge \delta$ as $d_{TV}(T_1, T_2) \le \delta$. 
We now claim that for $T$ to be extreme when $a_2=\delta$, it is required that $b_2=\delta$ or $b_2=\delta-c_7$. Otherwise the perturbation matrix $E$ such that $\epsilon_{2,1}=-\epsilon_{2,2}=\epsilon$ results in $T\pm E \in \Pd$ for some $ 0< \epsilon < \min(\delta-b_2, b_2+c_7-\delta)$. (a) When $a_2=b_2=0$ it can be shown that $c_7=\delta$. Otherwise perturbation matrix $E$ such that $\epsilon_{1,1}=\epsilon_{2,1}=\epsilon_{3,1}=-\epsilon_{1,3}=-\epsilon_{3,3}=-\epsilon_{2,5}=-\epsilon_{3,5}=\epsilon$ results in  $T\pm E \in \Pd$. (b) When $a_2=\delta$ and $b_2 = \delta - c_7$. 
$T$ matrix is therefore  of the form as shown below:
\[
\begin{bmatrix}
1-2\delta+c_7 & \delta & \delta-c_7 & 0 & 0 & 0 & 0\\
1-2\delta+2c_7 & \delta-c_7 &  0& 0 & \delta-c_7 & 0 & 0 \\
1-2\delta+c_7 & 0 & \delta-c_7  & 0 & \delta-c_7 & 0 & c_7 
\end{bmatrix}.
\]
The perturbation matrix $E$ such that $\epsilon_{1,1}=-\epsilon_{1,3}=-\epsilon_{2,2}=-\epsilon_{2,5}=\epsilon_{3,1}=-\epsilon_{3,3}=-\epsilon_{3,5}=\epsilon_{3,7}=\epsilon$ and $\epsilon_{2,1}=2\epsilon$ results in $T \pm E \in \Pd$ if $c_7 \ne \delta$. (c) When $a_2=b_2=\delta$ the perturbation matrix defined in the case (a) where $a_2=b_2=0$ can be used to show that $T \pm E \in \Pd$ unless $c_7 =\delta$.

\een

\subsubsection{Case 6: $a_2=b_2, a_3=c_3$} In this case $T$ is of the form:
\[
T=\begin{bmatrix}
1-a_2-a_3 & a_2 & a_3 & 0 & 0 & 0 & 0\\
1-a_2-b_5 & a_2 &  0& 0 & b_5 & 0 & 0 \\
1-a_3-c_5-c_7 & 0 & a_3  & 0 & c_5 & 0 & c_7 
\end{bmatrix}
\]
The $d_{TV}$ calculation between rows 1-2 and 1-3 gives max$\{a_3,b_5\} = \delta$
and max $\{a_2, c_5+c_7\} = \delta$. Note that $a_3+c_7 \le \delta$ and since $c_7 > 0$ it forces $a_3 < \delta$. Therefore, $b_5=\delta$ for rows 1 and 2 to be tight. 
\ben
\item $b_5 = a_2 = \delta$
The matrix structure is 
\[
\begin{bmatrix}
1-\delta-a_3&  \delta&  a_3 & 0 & 0 \\
1-2\delta &   \delta& 0& \delta & 0  \\
1-a_3-c_5-c_7& 0 & a_3&c_5  & c_7 
\end{bmatrix}
\]
Since $c_7 >0 $ and $c_5 + c_7 \le \delta$ it follows that $c_5 < \delta$. But that contradicts that the $d_{TV}(T_2, T_3) \le \delta$. Hence $T$ cannot be extreme in this case.

\item $b_5 = c_5+c_7=\delta$
The matrix structure is 
\[
\begin{bmatrix}
1-a_2-a_3&  a_2&  a_3 & 0 & 0 \\
1-a_2-\delta & a_2  & & 0& \delta & 0  \\
1-a_3-\delta& 0 & a_3& 0 &c_5  & \delta-c_5 
\end{bmatrix}
\]
Note that $a_2, a_3, c_5 < \delta$ as $a_3+c_7 \le \delta$, $c_5+c_7 \le \delta$,  $d_{TV}(T_2, T_3)\le \delta$ and $c_7 > 0$. Either $a_3=0$ or $a_3=c_5$. Otherwise perturbation matrix $\epsilon_{1,1}=\epsilon_{3,1}=-\epsilon_{1,3}=-\epsilon_{3,3}=\epsilon$ results in $T \pm E \in \Pd$. Similarly it can be shown that $a_2 = 0$ or $a_2 = c_5$. In all these cases the matrix $T$ reduces to a matrix with entries that are function of $c_5$. It can be shown that unless $c_5=0$ there is a perturbation matrix $E$ that results in $T \pm E \in \Pd$. This perturbation matrix can be constructed by placing $\epsilon$'s in the locations where $+c_5$ appears and $-\epsilon$'s in the locations where $-c_5$ appears.
\een
\end{appendices}

\end{document}